\documentclass[prd,twocolumn,showpacs,letterpaper,nofootinbib]{revtex4}
\usepackage{amssymb,amsmath}
\usepackage{color}
\usepackage{graphicx}

\newcommand{\bn}{\hat{\mathbf{n}}}
\newcommand{\br}{\mathbf{r}}
\newcommand{\bh}{\mathbf{H}}
\newcommand{\bk}{\mathbf{k}}

\newcommand{\bm}{\mathbf{m}}

\newcommand{\mypip}{\frac{1}{(2\pi)^{3/2}}\int d^3\bk e^{i\bk\cdot\mathbf{x}'}}
\newcommand{\beq}{\begin{equation}}
\newcommand{\eeq}{\end{equation}}
\newcommand{\bea}{\begin{eqnarray}}
\newcommand{\eea}{\end{eqnarray}}
\newcommand{\ba}{\begin{array}}
\newcommand{\ea}{\end{array}}

\newcommand{\ApJ}{Astrophys. J.}
\newcommand{\PRL}{Phys. Rev. Lett.}
\newcommand{\PRD}{Phys. Rev. D}

\newlength{\sizeonefig}
\newlength{\sizetwofig}
\begin{document}
\title{Weak Lensing of the Cosmic Microwave Background by Foreground Gravitational Waves}
\author{Chao Li$^1$}, \author{Asantha Cooray$^{1,2}$}
\affiliation{ $^1$ Theoretical Astrophysics, California Institute
of Technology, Mail Code 103-33 Pasadena, California 91125 \\
$^2$Center for Cosmology, Department of Physics and Astronomy, 4129  Frederick Reines Hall,
University of California,   Irvine, CA 92697}

\date{\today}

\begin{abstract}

Weak lensing distortion of the background cosmic microwave
background (CMB) temperature and polarization patterns by the
foreground density fluctuations is well studied in the literature.
We discuss the gravitational lensing modification to CMB
anisotropies and polarization by a stochastic background of
primordial gravitational waves between us and the last scattering
surface. While density fluctuations perturb CMB photons via
gradient-type deflections only, foreground gravitational waves
distort CMB anisotropies via both gradient- and curl-type
displacements. The latter is a rotation of background images,
while the former is related to the lensing convergence. For a
primordial background of inflationary gravitational waves, with an
amplitude corresponding to a tensor-to-scalar ratio below the
current upper limit of $\sim$ 0.3, the resulting modifications to
the angular power spectra of CMB temperature anisotropy and
polarization are below the cosmic variance limit.
At tens of arcminute angular scales and below,
these corrections, however, are above the level at which systematics must be controlled
in all-sky anisotropy and polarization maps with no instrumental noise and other secondary and foreground signals.

\end{abstract}

\maketitle

\section{Introduction}
The weak lensing of cosmic microwave background (CMB) anisotropies
and polarization by intervening mass fluctuations, or scalar
perturbations, is now well studied in the literature
\cite{lensing,Hu00}, with a significant effort spent on improving
the accuracy of analytical and numerical calculations (see, recent
review in \cite{Cha}). The non-Gaussian pattern of CMB
anisotropies and polarization created by non-linear mapping
associated with lensing angular deflections aids the extraction of
certain statistical properties of the foreground mass distribution
\cite{HuOka02}. Weak lensing deflections by intervening mass also
{\it leak} CMB polarization power in the E-mode to the B-mode
\cite{Zaldarriaga:1998ar}. This lensing B-mode signal presents a
significant confusion when searching for primordial gravitational
wave signatures in the CMB polarization \cite{KamKosSte97}. The
lensing reconstruction techniques discussed in the literature,
however, allow the possibility to ``clean'' CMB polarization maps
and to search for a background of inflationary gravitational waves
with an energy scale as low as $10^{15}$ GeV \cite{KesCooKam02}.

Similar to gravitational lensing by density perturbations, if
there is a background of gravitational waves in the foreground,
then one would expect metric perturbations associated with these
waves to distort and gravitationally lens background images
\cite{Jaffe}. While the lensing deflections by the density field
can be  written as the gradient of the projected gravitational
potential, lensing displacements due to gravitational waves can be
decomposed to both a gradient and a curl-like component
\cite{stebbins,DRS03,CooKamCal05}. In these two components,
gradient-type displacements are related to the lensing
convergence, while curl-type displacements are related to the
image rotation, though both types of displacements lead to image shear.
 While linear density perturbations do not produce rotations, second order corrections to weak lensing by scalar
perturbations, such as due to the coupling of two lenses along the
line of sight, can produce rotational modes \cite{CooHu02}.

While the study of CMB lensing by foreground density fluctuations
is now well developed \cite{Cha}, the discussion of CMB lensing by
foreground gravitational waves is limited.  In the context of
large-scale structure weak lensing surveys with galaxy shapes
\cite{Gallens}, the rotational power spectrum of background galaxy
images when lensed by primordial gravitational waves in the
foreground is discussed in Ref.~\cite{DRS03}. In the context of
lensing reconstruction with CMB temperature and polarization maps,
the curl component of the displacement field can be used to
monitor systematics \cite{CooKamCal05}, though lensing by
gravitational waves will leave a non-zero contribution to the curl
component.

Here, we extend the calculation in Ref.~\cite{DRS03} and study
both the curl- and the gradient-modes of the deflection field from
primordial gravitational waves that intervene CMB photons
propagating from the last scattering surface. Our calculations are
both useful and important given the increasing interest on, and
plans for, high sensitivity CMB anisotropy and polarization
measurements, including a potential space-based mission after
Planck, called CMBpol in the future.  Such an experiment is
expected to study polarization B-modes in exquisite detail and it
is important to understand potentially interesting secondary
signals beyond those that are routinely mentioned in the
literature. Based on the calculations presented here,
unfortunately,  we find that gravitational lensing of CMB by a
background of primordial gravitational waves from inflation, with
an amplitude below the current tensor-to-scalar ratio upper limit
of 0.3, will produce an undetectable modification to anisotropy
and polarization power spectra. Moreover, since the corrections
are below the cosmic variance level, it is unlikely that one needs
to account for these secondary corrections when making precise
cosmological measurements.

This paper is organized as follows. In Section~\ref{sec2}, we
discuss lensing by foreground gravitational waves by discussing
both the gradient and curl components of the displacement field.
Section~\ref{sec3} presents expressions for the weak lensing
correction to the CMB anisotropy and polarization power spectra.
We conclude with a discussion of our results in
Section~\ref{sec4}.

\begin{figure}[t]
\includegraphics[scale=0.4,angle=-90]{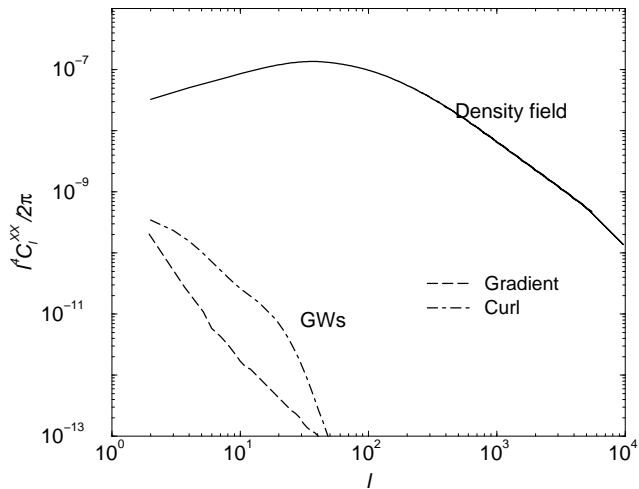}
\caption{Lensing-deflection power spectra. Here, we show the gradient
component from density perturbations (top curve), the curl (dot-dashed line) and gradient (dashed line)
components from foreground inflationary gravitational waves.  We have taken a background of gravitational waves with
an amplitude for the power spectrum corresponding to roughly
a tensor-to-scalar ratio of 0.3 or a Hubble parameter during inflation of $2 \times 10^{14}$ GeV.}
\label{spectra}
\end{figure}

\section{The spectrum of expansion and rotation}
\label{sec2}

To establish the lensing correction to CMB anisotropy and
polarization maps by foreground gravitational waves, we first need
to calculate the photon displacement on the spherical sky by
gravitational waves in the foreground. We make use of synchronous
coordinates and take the metric of a Friedman-Robertson-Walker
cosmological model as
\begin{equation}
g_{\mu \nu}=a^2\left( \begin{matrix}-1 & 0 \cr
                                0 & {\bf I}+{\bf H} \end{matrix} \right) \, ,
\end{equation}
where the scale factor is $a(\eta)$, and the conformal time is
denoted by $\eta$ with the value today of $\eta_0$. Here, ${\bf
H}$ is the transverse $(\nabla \cdot {\bf H}=0)$, symmetric $({\bf
H}={\bf H}^{\rm T})$,
 and traceless $({\rm Tr}\;{\bf H}=0)$ tensor
 metric perturbation associated with gravitational waves, while ${\bf I}$ is the identity matrix.
The photon propagation is governed by the geodesic equation \bea
\frac{d^2x^\mu}{d\lambda^2}+\Gamma^{\mu}_{\alpha\beta}
\frac{dx^\alpha}{d\lambda}\frac{dx^\beta}{d\lambda}=0.\eea By
changing variables to the conformal time
 $\eta$ from the affine parameter $\lambda$, the geodesic equation
 in the presence of foreground gravitational waves
 is \cite{DRS03}
\begin{equation}
\ddot{\bf r}= \frac{1}{2}\left(\dot{{\bf r}} \cdot \dot{\bf H} \cdot \dot{\bf r}\right)
\dot{\bf r}-\left({\bf I}+{\bf H}\right)^{-1} \cdot
\left[ \dot{\bf r} \cdot \frac{d}{d\eta}{\bf H} -
\frac{1}{2} \nabla_H \left(\dot{\bf r} \cdot {\bf H} \cdot \dot{\bf r}\right)\right] \, ,
\end{equation}
where, to simplify the notation, we have not written out the explicit dependence
of $\eta$ in each of these terms. In  here,
the over-dot represents the derivative with respect to the conformal time.
The full derivative $d/d\eta$ can be
separated to $\partial/\partial \eta + \dot{\bf r} \cdot \nabla$.
Here, and throughout, $\nabla_H$ denotes the gradient that applies only to
the metric perturbation ${\bf H}$;
when not subscripted with $H$, the gradient should be interpreted as the one that applies to all terms, including
the line of sight directional vector ${\bf n}$.

As gravitational fluctuation is very weak today, hereafter we
neglect $\mathbf{H}(\eta_0)$, and choose the initial conditions of
the trajectory to be \bea \mathbf{r}(\eta_0)=0\ ,\ \ \
\dot{\br}(\eta_0)=-\bn \,  \eea we find that the general
displacement on the celestial sphere induced by primordial
gravitational waves is \bea &&\mathbf{r}(\eta)=\bn(\eta_0-\eta)
-\int_{\eta}^{\eta_0}d\eta'\times\\
&&\left(\mathbf{H}\cdot\bn+\frac{1}{2}(\eta'-\eta)[(\bn\cdot
\dot{\mathbf{H}}\cdot\bn)\bn-\nabla_H(\bn\cdot{\mathbf{H}}\cdot
\bn)]\right)_{(\eta',\mathbf{x}')}\nonumber \label{eqnr} ,\eea
where $\mathbf{x}'=(\eta_0-\eta')\bn$. Similar to
Ref.~\cite{DRS03}, we have evaluated the trajectory on the
unperturbed path following the so-called Born approximation. One
could potentially evaluate corrections to this approximation in
terms of a perturbative correction to the path length, but these
would be at the second order in metric perturbations and will be
ignored.

The total displacement can be separated to a part along the line
of sight and a part perpendicular to it. The radial displacement
leads to a time-delay effect, similar to the lensing time-delay
associated with foreground potentials \cite{timedelay}. This
time-delay couples to the radial gradient of the CMB, due to
finite extent of the last scattering surface, while the angular
displacement couples to the angular gradient. As discussed in
Ref.~\cite{HuCoo01}, the overall correction to CMB anisotropy
spectra from the time-delay effect is subdominant since the spatial
gradient of the CMB is, at least, two orders of magnitude smaller
compared to the angular gradient. There are also geometrical cancellations that make
the time-delay effect smaller relative to angular deflections. Thus, we ignore the radial
displacement and only consider the transverse component related to
gravitational lensing angular deflections.

Using the transverse displacement, the angular deflection
projected on the spherical sky is $\vec{\Delta} = [{\bf r}(\eta) -
({\hat{\bf n}} \cdot {\bf r})\hat{\bf n}]/(\eta_0-\eta)$. These
two-dimensional displacements can be related to usual quantities
in gravitational lensing with the convergence and the rotation
defined as \cite{stebbins}
 \bea\kappa\equiv-\frac{\Delta^a{}_{:a}}{2},\ \quad {\rm and} \quad \
\omega\equiv\frac{(\Delta_a\epsilon^{ab})_{:b}}{2} \, ,\eea
respectively.

Similarly, we note that the general displacement on the celestial sphere can be decomposed to two components,
 \bea \Delta_a =-\sum_{lm}(h_{lm}^\oplus
Y_{lm:a}+h_{lm}^\otimes Y_{lm:b}\epsilon^b{}_a),
 \label{displacements}\eea
 where
 \bea h^\oplus_{lm}&=&\frac{1}{l(l+1)}\int d \bn
 Y_{lm}^*\Delta_a{}^{:a}=-\frac{2}{l(l+1)}\int d\bn Y_{lm}^*\kappa\nonumber\\
 h^\otimes_{lm}&=&\frac{1}{l(l+1)}\int d \bn
 Y_{lm}^*\Delta_a{}^{:b}\epsilon^a{}_{b}=\frac{2}{l(l+1)}\int d\bn
 Y_{lm}^*\omega, \nonumber \\
\eea where $\oplus$ and $\otimes$ denote the gradient- and
curl-type deflections, respectively. For simplicity, we have
dropped the dependence on the directional vector parameterized by
$\hat{\bf n}$. The lensing by foreground density perturbations to
the first-order only leads to a gradient-like  displacement, while
both components are generated when lensed by gravitational waves.
We now calculate both the convergence and the rotational spectrum
of the displacement field due to foreground stochastic
gravitational waves.

\subsection{Gradient Spectrum}

Gradient deflections are associated with the expansion and, as
defined above, can be described in terms of the convergence:
$\kappa(\bn)\equiv-\Delta^a{}_{:a}/2$. {\color{black} In general
the transverse divergence of a vector $\vec{A}$ can be rewritten
as\bea & &\frac{1}{\sin\theta}\frac{\partial
}{\partial\theta}(\sin\theta
A_\theta)+\frac{1}{\sin\theta}\frac{\partial A_\phi}{\partial
\phi}\nonumber\\&=&r\left(\nabla\cdot
\vec{A}-(\bn\cdot\nabla)(\bn\cdot \vec{A})-\frac{2\bn\cdot
\vec{A}}{r}\right),\eea where $r=\eta_0-\eta'$.} When substituting
the form of $\mathbf{r}(\eta)$ from equation~(5), the gradient
terms here lead to terms that are due to $\nabla \cdot {\bn}$ and
$\nabla \cdot {\bf H}$. We first consider the former and making
use of the fact that $\partial_i \hat{n}_j = (\delta_{ij} -
\hat{n}_i \hat{n}_j)/(\eta_0-\eta')$,
 we separate contributions to convergence to
two components and write $\kappa(\bn) = \kappa_1+\kappa_2$ as \bea
\kappa_1&=&-\frac{1}{2}\int_{\eta_s}^{\eta_0}
d\eta'{\color{black}\left(\frac{\eta_0-\eta'}{\eta_0-\eta_s}\right)}(\bn\cdot\nabla_H)
(\bn\cdot\bh\cdot\bn)\nonumber\\&
&-\frac{3}{2(\eta_0-\eta_s)}\int_{\eta_s}^{\eta_0}d\eta'
(\bn\cdot\bh\cdot\bn),\nonumber\\
\kappa_2&=&-\frac{1}{4}\int_{\eta_s}^{\eta_0}d\eta'{\color{black}
\left(\frac{\eta_0-\eta'}{\eta_0-\eta_s}\right)}
(\eta'-\eta_s)\nabla_H^2(\bn\cdot\bh\cdot\bn) \nonumber\\
& &+\frac{1}{4}\int_{\eta_s}^{\eta_0}d\eta'
{\color{black}\left(\frac{\eta_0-\eta'}{\eta_0-\eta_s}\right)}(\eta'-\eta_s)
(\bn\cdot\nabla_H)^2(\bn\cdot\bh\cdot\bn) \nonumber\\&
&+\frac{1}{\eta_0-\eta_s}\int_{\eta_s}^{\eta_0}d\eta'{(\eta'-\eta_s)}(\bn\cdot
\nabla_H)(\bn\cdot\bh\cdot\bn) \label{eqnk} \, ,
 \eea
where we have explicitly simplified the calculation by including terms associated with $\nabla \cdot \bn$.

Note that we have also replaced $\eta \rightarrow \eta_s$ corresponding to the
conformal time at the last scattering surface of CMB. To simplify,
we decompose the metric perturbation into the Fourier component,
\bea \mathbf{H}(\mathbf{x},\eta)=\frac{1}{(2\pi)^{3/2}}\int d^3
\mathbf{k} e^{i\mathbf{k}\cdot \mathbf{x}}T(k,\eta)\sum_{j=1}^2
H_j(\mathbf{k})\mathbf{e}_{j}(\mathbf{k}) \, ,\eea where we have
introduced the gravitational wave transfer function that describes
the time evolution of the metric perturbation with ${\bf
H}_j(\bk,\eta) = T(k,\eta)  H_j(\bk) \mathbf{e}_j(\bk)$.

The terms in equation~(\ref{eqnk}) can be simplified as \bea
\kappa_1&=&-\frac{1}{2}\int_{\eta_s}^{\eta_0}d\eta'\mypip
T(k,\eta')\nonumber\\
&&
{\color{black}\left(\frac{\eta_0-\eta'}{\eta_0-\eta_s}\right)}\times
\sum_j  H_{j}(\bk)[\bn\cdot\mathbf{e}_j(\bk)\cdot\bn](i\bk\cdot\bn)\, , \nonumber\\
& &-\frac{3}{2(\eta_0-\eta_s)}\int_{\eta_s}^{\eta_0}d\eta'
\mypip\nonumber\\
& &\times T(k,\eta')\sum_{j}H_j(\bk)[\bn\cdot\mathbf{e}_j(\bk)\cdot\bn]\nonumber\\
\kappa_2&=&\int_{\eta_s}^{\eta_0}d\eta'\frac{\eta'-\eta_s}{4}\mypip
T(k,\eta')\nonumber\\
&&
{\color{black}\left(\frac{\eta_0-\eta'}{\eta_0-\eta_s}\right)}\times
\sum_j {\bf
H}_{j}(\bk)[\bk^2-(\bn\cdot\bk)^2][\bn\cdot\mathbf{e}_j(\bk)\cdot\bn]\nonumber\\
& &
+\frac{1}{\eta_0-\eta_s}\int_{\eta_s}^{\eta_0}d\eta'(\eta'-\eta_s)\mypip\nonumber\\
& &\times T(k,\eta')\sum_j
H_{j}(\bk)[\bn\cdot\mathbf{e}_j(\bk)\cdot \bn](i\bk\cdot \bn).
\label{eqnkappa} \eea Here, ${\bf x}'=(\eta_0-\eta'){\hat {\bf
n}}$ and $\mathbf{e}_j$ represents the symmetric, traceless
polarization tensor that obeys ${\rm Tr}\; [\mathbf{e}_j(\bk)
\cdot \mathbf{e}_k(\bk)]=2 \delta_{jk}$ and ${\bf k} \cdot
\mathbf{e}_j(\bk)=0$ with $j$ in equation~(\ref{eqnkappa}) summing
over the two linear polarization states.

Since gravitational waves trace the wave equation with \bea
\ddot{{\bf H}} - \nabla^2 {\bf H} + 2 \frac{\dot{a}}{a} {\bf H}=
16 \pi G a^2 {\bf P} \, \eea where ${\bf P}$ is the tensor part of
the anisotropic stress, say from neutrinos
 (see, Ref.~\cite{Prit} for details); The term in the right hand side acts as a damping term for the evolution of gravitational waves
and is important for modes that enter the horizon before
matter-radiation equality, with a smaller correction for modes
enter horizon after matter-radiation equality. Since these
corrections are not more than 30\%, while the amplitude of the
gravitational wave background is uncertain to more than orders of
magnitude, we ignore such subtleties here assuming no anisotropic
stress; for a cosmological model dominated by matter,
 in Fourier space, one can write the evolution of ${\bf H}$ in
 terms of the transfer function as $T(k,\eta)=3j_1(k\eta)/(k\eta)$.

We define the power spectrum of metric perturbations as
\begin{equation}
\langle H_i(\bk) H_j^*(\bk) \rangle = (2\pi)^3 P_T(k) \delta_{ij}
\delta^{(3)}(\bk-\bk') \, ,
\end{equation}
where we assume isotropy and equal density of gravitational waves in the two polarization states $i$ and $j$.
Following Ref.~\cite{DRS03}, we normalize the power spectrum to the Hubble parameter during inflation and take
\begin{equation}
P_T(k) = \frac{8\pi}{(2\pi)^3} \left(\frac{H_I}{M_{\rm Planck}}\right)^2 k^{-3} \, .
\label{eqnpk}
\end{equation}

Using this three-dimensional power spectrum for metric
perturbations, the angular power spectrum of gradient-type
deflections is
\bea
C_l^{h^\oplus}&=&\frac{1}{2l+1}\sum_{m=-l}^{m=l}\langle
|h^{\oplus}_{lm}|^2\rangle \\
&=&\frac{4}{(2l+1)l^2(l+1)^2}\sum_{m=-l}^{m=l}\langle
|\int d\bn Y_{lm}^*(\bn)\kappa(\bn)|^2\rangle\nonumber\\
&=&\frac{\pi}{l^2(l+1)^2}\frac{(l+2)!}{(l-2)!}\int d^3\bk
P_{T}(k)|T_1+T_2+T_m^\oplus|^2, \nonumber
\label{partconvergence}\eea where the terms are again \bea
T_1&=&-k\int_{\eta_s}^{\eta_0}d\eta'
{\color{black}\left(\frac{\eta_0-\eta'}{\eta_0-\eta_s}\right)}T(k,\eta')\nonumber\\
& &\quad \times
\left[\partial_x(x^{-2}j_l(x))|_{x=k(\eta_0-\eta')}
\right]\nonumber\\
& &-\frac{3}{\eta_0-\eta_s}\int^{\eta_0}_{\eta_s}d\eta'
T(k,\eta')(x^{-2}j_l(x))|_{x=k(\eta_0-\eta')},
\nonumber\\
T_2&=&\frac{k^2}{2}\int_{\eta_s}^{\eta_0}d\eta'
{\color{black}\left(\frac{\eta_0-\eta'}{\eta_0-\eta_s}\right)}
(\eta'-\eta_s)T(k,\eta') \nonumber \\
&& \quad \quad  \times
\left[(1+\partial_x^2)(x^{-2}j_l(x))|_{x=k(\eta_0-\eta')}\right]\nonumber\\
&
&+\frac{2k}{\eta_0-\eta_s}\int_{\eta_s}^{\eta_0}d\eta'(\eta'-\eta_s)T(k,\eta')\nonumber\\
& & \quad \quad \times
\left[\partial_x(x^{-2}j_l(x))|_{x=k(\eta_0-\eta')} \right]. \eea
Note that the convergence power spectrum due to gravitational wave
deflections is
 $C_l^{\kappa\kappa}=l^2(l+1)^2C_l^{h ^\oplus}/4$.

In equation.~(16), we have also introduced an additional
correction to the gradient-type deflection spectrum with the term
$T_m^\oplus$. As discussed in \cite{DRS03}, this is associated
with the gradient-type deflection pattern or convergence related
to shearing of the last  scattering surface by gravitational waves
present at that surface.  To calculate this correction, we first
evaluate the displacement vector at the last scattering surface
\bea \br_m =-\frac{\bh}{2}\cdot \bn (\eta_0-\eta_s) \, , \eea and
then project this displacement vector to obtain the transverse
displacement vector of  \bea \Delta_m=\frac{\br_m-(\br_m\cdot
\bn)\bn}{\eta_0-\eta_s}\, . \label{eqn:deltarm}\eea This vector
can be decomposed to the gradient- and curl-types deflections.
Making use of the fact that the convergence is
$\kappa(\bn)\equiv-\Delta_m^a{}_{:a}/2$ and taking the Fourier
transforms, we write the required term in the equation.~(16) as
\bea
T_m^\oplus&=&-\frac{k}{2}(\eta_0-\eta_s)T(k,\eta_s)\left[\partial_x(x^{-2}j_l(x))|_{x=k(\eta_0-\eta_s)}
\right]\nonumber\\ &
&-T(k,\eta_s)(x^{-2}j_l(x))|_{x=k(\eta_0-\eta_s)} \, . \eea This
completes the calculation of gradient-type deflection power
spectrum by taking into account the lensing by intervening
gravitational waves between the last scattering surface and the
observer and the metric-shear correction related to gravitational
waves present at the last scattering surface from which photons
propagate.

\subsection{Curl spectrum}

Rotation is defined to be $\omega(\bn) \equiv \frac{1}{2}(\Delta_a\epsilon^{ab})_{:b}$, which leads to
\begin{equation}
\omega(\bn)\equiv-\frac{1}{2}\bn\cdot(\nabla\times\br(\bn,\eta_s))
\,.
\end{equation}
Equation~(\ref{eqnr}) gives \bea \omega&=&
\frac{1}{2}\int_{\eta_s}^{\eta_0}d\eta' \left[\bn\cdot(\nabla
\times {\bf H}) \cdot \bn\right] \, ,\eea since $\nabla \times
\bn$ =0. Here we define the curl of the second rank tensor $\bf H$
by $(\nabla \times {\bf H})_{il} = \epsilon_{ijk}
\partial_j H_{kl}$.

 The curl-type deflection spectrum from this term is \cite{DRS03},
\bea
C_l^{h^\otimes}&=&\frac{1}{2l+1}\sum_{m=-l}^{m=l}\langle
|h^{\otimes}_{lm}|^2\rangle  \nonumber \\
&=&
\frac{4}{(2l+1)l^2(l+1)^2}\sum_{m=-l}^{m=l}\langle
|\int d\bn Y_{lm}^*(\bn)\omega(\bn)|^2\rangle\nonumber\\
&=&\frac{\pi}{l^2(l+1)^2} \frac{(l+2)!}{(l-2)!} \int d^3\bk
P_{T}(k)|T_3+T_m^\otimes|^2, \eea where \bea
T_3&=&2k\int_{\eta_s}^{\eta_0}d\eta'T(k,\eta')(x^{-2}j_l(x))|_{x=k(\eta_0-\eta')}\,
, \eea Again, one can write the power spectrum of rotation as
$C_l^{\omega \omega}=l^2(l+1)^2C_l^{h^\otimes}/4$.

Here also we include the correction to the rotational spectrum due to metric
perturbations  at the last scattering surface. Similar to convergence, by following the same procedure as before
but taking  $1/2([\Delta_m]_a\epsilon^{ab})_{:b}$ of Equation~(\ref{eqn:deltarm}),
we get
\bea
T_m^\otimes=k(\eta_0-\eta_s)T(k,\eta_s)(x^{-2}j_l(x))|_{x=k(\eta_0-\eta_s)}.
\eea

For comparison, we note that density fluctuations along the line of sight lead to gradient-type
deflections only. The resulting contributions are described in terms of
the angular power spectrum of projected potential $C_l^{\phi
\phi}$ that is well studied in the literature \cite{Hu00}. We do
not repeat those derivations here, but will provide a comparison
of lensing under gravitational waves and lensing by mass in the
discussion later.

\begin{figure*}[t]
\includegraphics[scale=0.4,angle=-90]{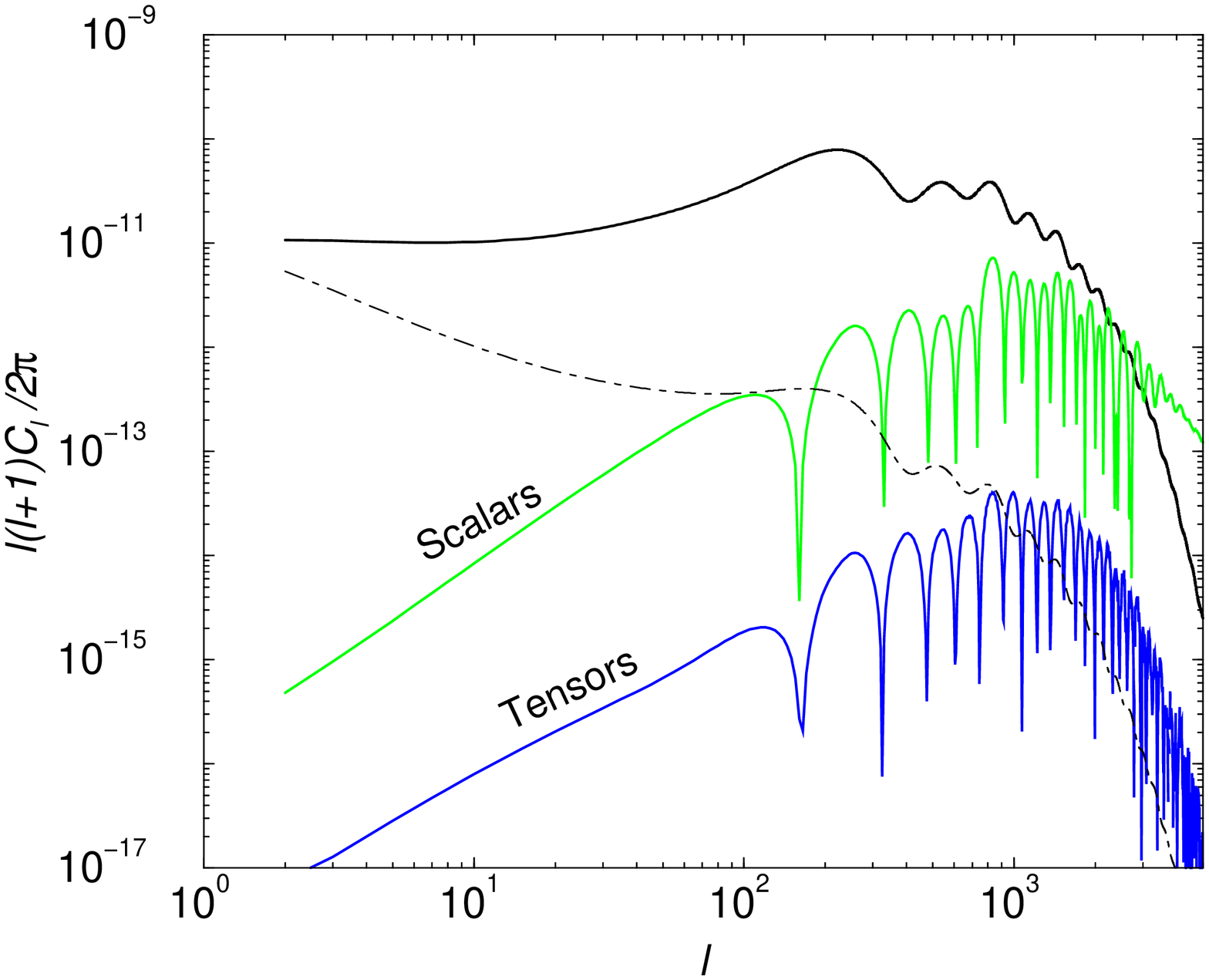} \hspace{1.0cm}
\includegraphics[scale=0.4,angle=-90]{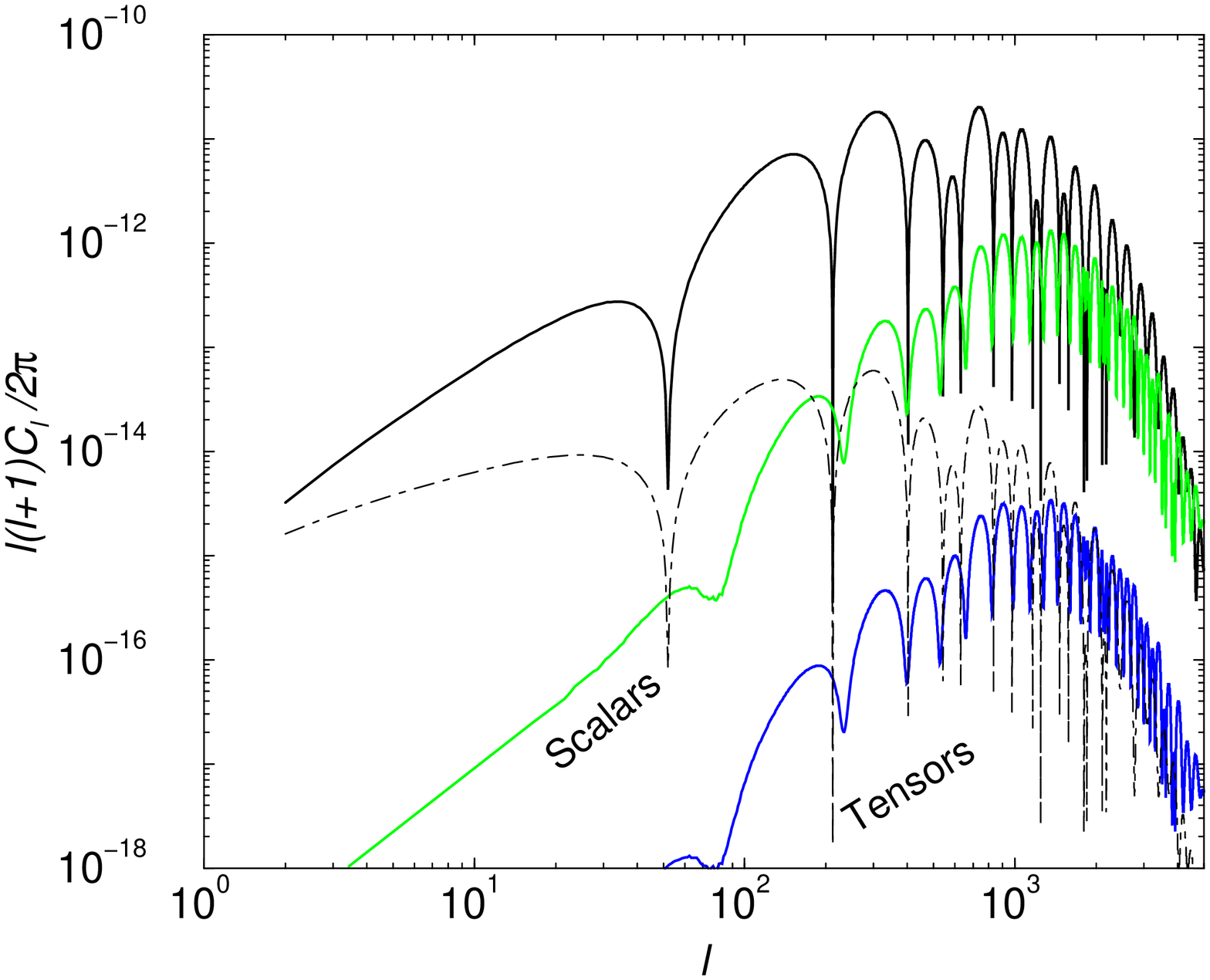}
\caption{The lensing modification to CMB power spectra for
density perturbations and for gravitational waves. {\it Left:} Temperature
fluctuations.  The top curve is the primordial power spectrum.
The middle curve is the secondary anisotropy contribution,
$|\tilde C_l - C_l|$ , to the temperature power spectrum from lensing by density
perturbations, and the lower curve is for lensing by
gravitational waves, assuming the maximum IGW background
consistent with current data with the deflection power spectra shown in Figure~1.
{\it Right:} Temperature-E polarization cross correlation, with curves following the
left panel. In both panels, thin long-dashed line is $C_l/l$ at each multipole
with $C_l$ related to the intrinsic anisotropy spectrum; while the cosmic variance is $C_l/\sqrt{l}$,
$C_l/l$ denotes the level at which one must control the systematics, if the effects resulting
systematics apply to a wide range of multipoles. The lensing by
density fluctuations cannot be ignored as the corrections are well above the cosmic variance limit
and will be detectable in upcoming anisotropy data. The lensing by foreground gravitational waves, however, are below the cosmic variance limit,
suggesting that they will remain undetectable, but above the systematic level when $l >10^3$.}
\label{cmb}
\end{figure*}

\section{Lensing of CMB by Gravitational Waves}
\label{sec3}

In this Section, we will discuss the analytical calculation related to how foreground gravitational waves modify the background
CMB temperature anisotropy and polarization patterns. Here, we will concentrate on the angular power spectra of CMB observables.
The lensing of CMB by foreground density fluctuations is formulated in Ref.~\cite{Hu00} and we follow the same procedure here.
Since deflection field power spectra peak at large angular scales, we present an analytical formulation as appropriate for the spherical sky.
The calculation we consider here, however, is perturbative and one expects important corrections beyond the first order in the deflection angle.
For foreground gravitational waves, such issues can be ignored as the overall modification to the anisotropy and polarization
spectra is small.

\subsection{Temperature Anisotropies}

Following Ref.~\cite{Hu00}, the lensed temperature field
$\tilde\theta$ can be expressed as \bea \tilde{\theta}
(\bn)=\theta(\bn+\mathbf{\Delta})=\theta(\bn)+\nabla^a\theta\cdot
\Delta_a+\frac{1}{2}\nabla^b\nabla^a\theta\cdot\Delta_a\Delta_b \, , \nonumber \\
\label{eqn:lenst} \eea where $\theta(\bn) $ is the unlensed
temperature fluctuation in the direction $\bn$. The temperature
field can be expanded to multipole moments such that $\theta(\bn)
= \sum_{lm} \theta_{lm} Y_{lm}(\bn)$. Taking the spherical
harmonic moment of equation~(\ref{eqn:lenst})  and using
equation~(\ref{displacements}), we find \bea &&
\tilde{\theta}_{lm}=\theta_{lm}+\int d\bn Y_{lm}^*
\left[\nabla^a\theta\cdot
\Delta_a+\frac{1}{2}\nabla^b\nabla^a\theta\cdot\Delta_a\Delta_b\right]
\nonumber\\
&&=\theta_{lm}-\sum_{l_1m_1l_2m_2}
\Big(I_{lml_1m_1l_2m_2}^\oplus\theta_{l_1m_1}h_{l_2m_2}^\oplus \nonumber \\
&& \quad +
I_{lml_1m_1l_2m_2}^\otimes\theta_{l_1m_1}h_{l_2m_2}^\otimes \Big)\nonumber\\
&&
+\frac{1}{2}\sum_{l_1m_1l_2m_2l_3m_3}\Big(J_{lml_1m_1l_2m_2l_3m_3}^\oplus
\theta_{l_1m_1}h_{l_2m_2}^\oplus h_{l_3m_3}^{\oplus*} \nonumber \\
&& \quad +
J_{lml_1m_1l_2m_2l_3m_3}^\otimes\theta_{l_1m_1}h_{l_2m_2}^
\otimes h_{l_3m_3}^{\otimes*}\Big), \eea
where  the integrals are
\begin{widetext}

\bea I_{lml_1m_1l_2m_2}^\oplus=&\int d\bn
Y_{lm}^*Y_{l_1m_1}^{:a}Y_{l_2m_2:a}\quad , \quad
I_{lml_1m_1l_2m_2}^\otimes&=\int  d\bn
Y_{lm}^*Y_{l_1m_1}^{:a}Y_{l_2m_2:b}\epsilon^b{}_a \nonumber\\
J_{lml_1m_1l_2m_2l_3m_3}^\oplus=&\int d\bn
Y_{lm}^*Y_{l_1m_1}^{:ab}Y_{l_2m_2:a}Y_{l_3m_3:b}^*\quad ,\quad
J_{lml_1m_1l_2m_2l_3m_3}^\otimes&=\int d\bn
Y_{lm}^*Y_{l_1m_1}^{:ab}Y_{l_2m_2:c}Y_{l_3m_3:d}^*\epsilon^c{}_a\epsilon^d{}_b
\, . \eea  The lensed temperature anisotropy power spectrum
spectrum is \bea C^{\tilde{\theta}}_l
=C_l^{\theta}+\sum_{l_1l_2}C_{l_1}^{\theta}\left(C_{l_2}^{h^{\oplus}}S_1^{\oplus}+
C_{l_2}^{h^{\otimes}}S_1^{\otimes}\right)  \quad
+C_l^{\theta}\sum_{l_1}
\left(C_{l_1}^{h^{\oplus}}S_2^{\oplus}+C_{l_1}^{h^{\otimes}}S_2^{\otimes}\right),\eea
\end{widetext}
where \bea S_1^\oplus &=&
\sum_{m_1m_2}|I_{lml_1m_1l_2m_2}^\oplus|^2 \\ S_1^\otimes
&=& \sum_{m_1m_2}|I_{lml_1m_1l_2m_2}^\otimes|^2 \nonumber\\
S_2^\oplus
&=&\frac{1}{2}\sum_{m_1}J_{lmlml_1m_1l_1m_1}^\oplus+{\rm
c.c.}\nonumber\\
S_2^\otimes&=&\frac{1}{2}\sum_{m_1}J_{lmlml_1m_1l_1m_1}^\otimes+{\rm
c.c.}\, , \nonumber \eea and ${\rm c.c.}$ is the complex
conjugate. The terms $S_1^\oplus$ and $S_2^\oplus$ are similar to
those involving lensing by foreground density perturbations
\cite{Hu00}. First, the integral $I_{lml_1m_1l_2m_2}^\oplus$ can
be simplified through integration by parts and noting $\nabla^2
Y_{lm} = -l(l+1)Y_{lm}$ and the general integral of three
spin-spherical harmonics over the sky: \bea && \int d\bn
({}_{s_1}Y_{l_1m_1}^*)({}_{s_2}Y_{l_2m_2})({}_{s_3}Y_{l_3m_3}) =
\nonumber \\
&& \quad (-1)^{m_1+s_1}
\sqrt{\frac{(2l_1+1)(2l_2+1)(2l_3+1)}{4\pi}} \nonumber \\
&& \quad \quad \times \begin{pmatrix}l_1&l_2&l_3\\
s_1&-s_2&-s_3\end{pmatrix}\begin{pmatrix}l_1&l_2&l_3\\-m_1&m_2&m_3\end{pmatrix}
\, \label{eqnylm} \eea when $s_1=s_2+s_3$. We note that under
parity inversion, ${}_{s}Y_{lm} \rightarrow (-1)^l {}_{-s}Y_{lm}$,
which is a useful property when we discuss lensing modifications
to the CMB polarization field.

\begin{figure}[t]
\includegraphics[scale=0.4,angle=-90]{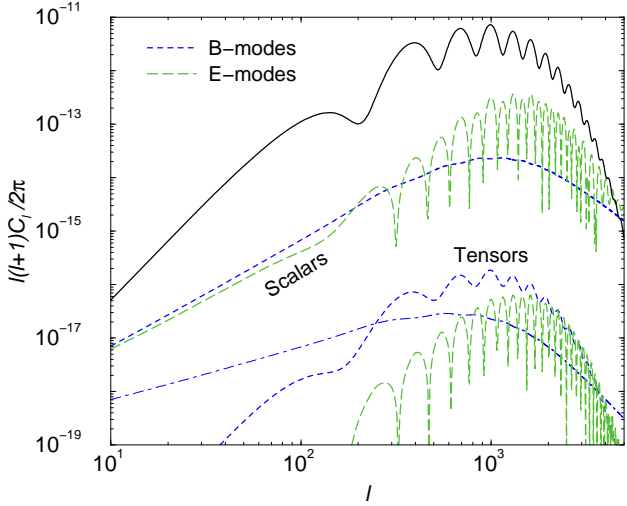}
\caption{The lensing modification to CMB polarization spectra for
density perturbations and for gravitational waves.  From top to
bottom, these curves are for primordial power spectrum in the
E-mode (solid line), lensing correction to E-mode from density
fluctuations (long -dashed line), lensing correction to B-mode
from density fluctuations (short-dashed line labeled ``Scalars''),
lensing correction to B-mode from gravitational waves
(short-dashed line labeled ``Tensors''), and the lensing
correction to E-mode from gravitational waves (long-dashed line).
For reference, the thin dot-dashed line is the systematic level of
the B-mode lensing power spectrum, $C_l/l$, from the density
field. While the corrections from lensing by foreground
gravitational waves is below the cosmic variance limit of the
lensing B-mode power spectrum and, again, undetectable in
anisotropy maps, they may become a source of systematic in all-sky
maps with no instrumental noise and other secondary signals and
foregrounds.} \label{cmbpol}
\end{figure}

With $s_i=0$ and noting that ${}_{0}Y_{lm}=Y_{lm}$,
\bea
&&I_{lml_1m_1l_2m_2}^\oplus=\frac{1}{2}[l_1(l_1+1)+l_2(l_2+1)-l(l+1)](-1)^{m} \nonumber \\
&&\times \sqrt{\frac{(2l+1)(2l_1+1)(2l_2+1)}{4\pi}}\nonumber\\
&&\times \begin{pmatrix}l&l_1&l_2\\
0&0&0\end{pmatrix}\begin{pmatrix}l&l_1&l_2\\-m&m_1&m_2\end{pmatrix}
\,. \eea Using the orthonormality relation of Wigner-3j symbols
\begin{equation}
\sum_{m_1m_2}
\begin{pmatrix}l_1&l_2&l_3\\
m_1&m_2&m_3\end{pmatrix}\begin{pmatrix}l_1&l_2&l_3\\m_1&m_2&m_3\end{pmatrix}
= \frac{1}{2l_3+1} \, , \label{eqnortho}
\end{equation}
we can write
\bea
S_1^\oplus=\frac{1}{2l+1}(F_{ll_1l_2}^\oplus)^2,\eea
with
\bea
&&F_{ll_1l_2}^\oplus=\frac{1}{2}\left[l_1(l_1+1)+l_2(l_2+1)-l(l+1)\right] \nonumber \\
&\times& \sqrt{\frac{(2l+1)(2l_1+1)(2l_2+1)}{4\pi}}\begin{pmatrix}
l&l_1&l_2\\0&0&0\end{pmatrix} \, .
\label{eqnfplus} \eea

Though tedious, the calculation related to  $S_1^\otimes$ can be
simplified using the gradient relation  for spherical harmonics by
raising and lowering of the spin \cite{Gol67}: \bea \nabla
Y_{lm}=\sqrt{\frac{l(l+1)}{2}}[{}_{1}Y_{lm}\mathbf{m}_+-{}_{-1}Y_{lm}\mathbf{m}_-],
\eea where \bea
\mathbf{m}_{\pm}=\frac{1}{\sqrt{2}}(e_{\hat{\theta}}\mp i
e_{\hat{\phi}}) \, .\eea Combining these derivatives with the
general integral in equation~(\ref{eqnylm}) leads to \bea
&&I_{lml_1m_1l_2m_2}^\otimes=-\frac{i}{2}
\sqrt{l_1(l_1+1)l_2(l_2+1)}
(-1)^{m} \nonumber \\
&\times& \sqrt{\frac{(2l+1)(2l_1+1)(2l_2+1)}{4\pi}}\nonumber\\
&\times&\begin{pmatrix}l&l_1&l_2\\-m&m_1&m_2\end{pmatrix}
\begin{pmatrix}l&l_1&l_2\\
0&-1&1\end{pmatrix}\left[1-(-1)^{l+l_2+l_2}\right] \, .
\eea

Again using the orthonormality relation in
equation~(\ref{eqnortho}),
\bea
S_1^\otimes=\frac{1}{2l+1}\left(F_{ll_1l_2}^\otimes\right)^2, \eea
with \bea
&&F_{ll_1l_2}^\otimes=\frac{1}{2}\sqrt{l_1(l_1+1)l_2(l_2+1)}
\begin{pmatrix}l&l_1&l_2\\ 0&-1&1\end{pmatrix} \\
&&\sqrt
{\frac{(2l+1)(2l_1+1)(2l_2+1)}{4\pi}} \left[1-(-1)^{l+l_1+l_2}\right] \, .
\nonumber
\label{eqnfcross}
\eea

To calculate  $S_2^\otimes$ and $S_2^\oplus$, we first note that
\bea \sum_{m_1}
Y_{l_1m_1:a}Y_{l_1m_1:b}^*=\frac{l_1(l_1+1)(2l_1+1)}{8\pi}g_{ab}
\eea where $g_{ab}$ is the usual metric of unit sphere,
\begin{equation}
g_{ab} =
\left( \begin{matrix} 1 & 0 \cr
                                0 & \sin^2\theta \end{matrix}\right) \, .
\end{equation}
These allow us to show that
\bea
S_2^\oplus &\equiv&
\frac{1}{2}\sum_{m_1}J^\oplus_{lmlml_1m_1l_1m_1}+{\rm c.c.} \nonumber \\
&=& -l(l+1)l_1(l_1+1)\frac{2l_1+1}{8\pi} \, .
\label{eqns}
\eea
Also $S_2^\otimes=S_2^\oplus$.
Finally, combining all expressions, we can write
\bea
C^{\tilde{\theta}}_l &=& C_l^{\theta}  - l(l+1)RC_l^{\theta}  \nonumber \\
&+& \sum_{l_1l_2}
\frac{C_{l_1}^{\theta}}{2l+1}\left[C_{l_2}^{h^{\oplus}}(F^\oplus_{ll_1l_2})^2
+ C_{l_2}^{h^{\otimes}}(F^\otimes_{ll_1l_2})^2\right]\, , \eea
where $F^\oplus_{ll_1l_2}$ and $F^\otimes_{ll_1l_2}$ are given in
equations~(\ref{eqnfplus}) and (40), respectively, and
\begin{equation}
R = \sum_{l_1} l_1(l_1+1) \frac{2l_1+1}{8\pi} \left[C_{l_1}^{h^{\oplus}}+C_{l_1}^{h^{\otimes}}\right] \, .
\label{eqnR}
\end{equation}
This expression is similar to that of equation~(62) of
Ref.\cite{Hu00} when $C_{l_1}^{h^{\otimes}}=0$ and
$C_{l_1}^{h^{\oplus}}$ is identified as the power spectrum of
projected lensing potentials due to intervening density
perturbations between us and the CMB.

\subsection{Polarization}

The lensing effect on CMB polarization can be described similar to
temperature anisotropies by making use of the remapping ${}_{\pm}
\tilde{X}(\bn) =  {}_{\pm}X(\bn) + \nabla_i \phi(\bn) \nabla^i {}_{\pm}X(\bn) +...+$,
where ${}_{\pm}X=Q \pm iU$, where we have simplified the notation by replacing the spin-dependent
gradients with a covariant derivative that acts on the spin-components of the
symmetric tensors that are traceless. Here, we have also ignored the rotation needed to align the polarization basis vectors between the lensed
and unlensed fields, but this rotation is unimportant when considering displacements along the lines of constant
azimuthal angles \cite{spin1}.  We refer the reader to Refs.~\cite{spin1,spin2} for details of our
shorthand notation and why it can be used for the lensing of the polarization pattern on the spherical sky.
While our notation here is simple and follows that of Ref.~\cite{Hu00},  the final result is the same from what one gets by
using the standard notation of differential geometry. This is due to the fact that the overlapping integrals  involving
spin harmonics that we will perform remain consistent with our simplified notation.

As is well known, the CMB polarization components form a spin-2 field and are
expanded in terms of the spin-weighted spherical harmonics such
that ${}_{\pm}X(\bn) = \sum_{lm} {}_{\pm}X_{lm} {}_{\pm
2}Y_{lm}(\bn)$. Instead of Stokes parameters, the more popular,
$E$ and $B$ modes are are given by ${}_{\pm} X_{lm} = E_{lm} \pm i
B_{lm}$. We will discuss lensing modifications to angular power
spectra of E- and B-modes as well as the cross-correlation between
E and $\theta$, while cross-correlation between E and B, and
between B and $\theta$ are ignored as these are zero through
parity arguments. Furthermore, the modifications to polarization
by lensing do not violate parity conservation.

Taking the spherical harmonic moment of the polarization field
under lensing, we write \bea {}_\pm\tilde{X}_{lm}&=&{}_\pm
X_{lm}-\\ \sum_{l_1m_1l_2m_2}&&\Big[{}_\pm X_{l_1m_1}\Big(
{}_{\pm 2}I^\oplus_{lml_1m_1l_2m_2}h^\oplus_{l_2m_2}\nonumber\\
 &&+ {}_{\pm 2} I^\otimes_{lml_1m_1l_2m_2}h^\otimes_{l_2m_2}\Big)\Big]
 \nonumber\\ +\frac{1}{2}\sum_{l_1m_1l_2m_2l_3m_3}&&\Big[{}_\pm
X_{l_1m_1}\Big({}_{\pm
2}J^\oplus_{lml_1m_1l_2m_2l_3m_3}h^\oplus_{l_2m_2}h^{\oplus
*}_{l_3m_3}  \nonumber\\ & &+ {}_{\pm
2}J^\otimes_{lml_1m_1l_2m_2l_3m_3}h^\otimes_{l_2m_2}h^{\otimes
*}_{l_3m_3}\Big)\Big],\nonumber \eea
  where \bea {}_{\pm 2}I_{lml_1m_1l_2m_2}^\oplus =&&\int d\bn
({}_{\pm
2}Y_{lm}^*)({}_{\pm 2}Y_{l_1m_1}^{:a})Y_{l_2m_2:a}\nonumber\\
{}_{\pm 2}J_{lml_1m_1l_2m_2l_3m_3}^\oplus =&&\int d\bn ({}_{\pm
2}Y_{lm}^*)({}_{\pm
2}Y_{l_1m_1}^{:ab})Y_{l_2m_2:a}Y_{l_3m_3:b}^*\nonumber\\
{}_{\pm 2}I_{lml_1m_1l_2m_2}^\otimes=&&\int  d\bn ({}_{\pm
2}Y_{lm}^*)({}_{\pm 2}Y_{l_1m_1}^{:a})Y_{l_2m_2:b}\epsilon^b{}_a
\nonumber\\ {}_{\pm 2}J_{lml_1m_1l_2m_2l_3m_3}^\otimes =&&\\
\int d\bn ({}_{\pm 2}Y_{lm}^*)&&({}_{\pm
2}Y_{l_1m_1}^{:ab})Y_{l_2m_2:c}Y_{l_3m_3:d}^*\epsilon^c{}_a\epsilon^d{}_b
\,. \nonumber \eea

After straightforward but tedious algebra, the lensed power
spectra of $\tilde{E}$-modes, $\tilde{B}$-modes, and the
cross-correlation between $\tilde{E}$-modes and $\tilde{\theta}$
are
\begin{widetext}
\bea
C_{l}^{\tilde{E}}&=&C_{l}^E+\frac{1}{2}\sum_{l_1l_2}\left[C_{l_1}^
{h^\oplus}{}_{2}S_1^\oplus+C_{l_1}^{h^\otimes}{}_{2}S^{\otimes}_1\right]\left[(C_{l_2}^E+C_{l_2}^B)+(-1)^L
(C_{l_2}^E-C_{l_2}^B)\right]+C_l^E \sum_{l_1} (C_{l_1}^{h^\oplus}{}_{2}S_2^\oplus+C_{l_1}^{h^\otimes}{}_{2}S_2^\otimes), \nonumber \\
C_{l}^{\tilde{B}}&=&C_{l}^B+\frac{1}{2}\sum_{l_1l_2}\left[C_{l_1}^{h^\oplus}{}_{2}S_1^\oplus+C_{l_1}
^{h^\otimes}{}_{2}S^{\otimes}_1\right]\left[(C_{l_2}^E+C_{l_2}^B)+(-1)^L(C_{l_2}^B-C_{l_2}^E)\right]+C_l^B \sum_{l_1}
(C_{l_1}^{h^\oplus}{}_{2}S_2^\oplus+C_{l_1}^{h^\otimes}{}_{2}S_2^\otimes) ,\nonumber \\
C_l^{\tilde{\theta}\tilde{E}}&=&C_l^{\theta E}+\frac{1}{2}\sum_{l_1l_2}(1+(-1)^L)(C_{l_1}^{h^\oplus}{}_{02}S_1^\oplus+C_{l_1}^
{h^\otimes}{}_{02}S_1^\otimes)C_{l_2}^{\theta E}+\frac{1}{4}C_l^{\theta
E}\sum_{l_1}\left(C_{l_1}^{h^\oplus}({}_{2}S_2^\oplus+S_2^\oplus)+C_{l_1}^{h^\otimes}
({}_{2}S_2^\otimes+S_2^\otimes)\right) \, ,
\eea
\end{widetext}
where $L=l+l_1+l_2$ and
\bea
{}_{2}S_1^\oplus&=&\sum_{m_1m_2}|{}_{\pm2}I_{lml_1m_1l_2m_2}^\oplus|^2, \\
{}_{2}S_1^\otimes &=&\sum_{m_1m_2}|{}_{\pm2}I_{lml_1m_1l_2m_2}^\otimes|^2,\nonumber\\
{}_{2}S_2^\oplus&=&\frac{1}{2}\sum_{m_1}{}_{\pm2}J_{lmlml_1m_1l_1m_1}^\oplus+{\rm c.c.,} \, \nonumber\\
{}_{2}S_2^\otimes&=&\frac{1}{2}\sum_{m_1}{}_{\pm2}J_{lmlml_1m_1l_1m_1}^\otimes+{\rm c.c.} , \nonumber\\
{}_{02}S_1^\oplus&=&\sum_{m_1m_2}(I_{lml_1m_1l_2m_2}^\oplus \;  {}_{+2}I_{lml_1m_1l_2m_2}^\oplus), \nonumber\\
{}_{02}S_1^\otimes&=&\sum_{m_1m_2}(I_{lml_1m_1l_2m_2}^\otimes \;
{}_{+2}I_{lml_1m_1l_2m_2}^\otimes). \label{eqnsum}
\nonumber \eea

To simplify terms in equation~(49), we again make use of the
integral relations outlined earlier when describing lensing of
temperature anisotropies. In the case of polarization, these
relations need to be generalized for integrals over spin-weighted
spherical harmonics. First, the integral related to the gradient
spectra is straightforward. Making use of the fact that
 $\nabla^2 {}_{\pm 2} Y_{lm} = [-l(l+1)+4] {}_{\pm 2}Y_{lm}$ and using equation~(\ref{eqnylm}), we find
\bea
&&{}_{2}S_1^\oplus=\frac{1}{2l+1}|{}_{2}F_{ll_1l_2}^\oplus|^2,\nonumber\\
&&{}_{2}F_{ll_1l_2}^\oplus=\frac{1}{2}[l_1(l_1+1)+l_2(l_2+1)-l(l+1)] \nonumber \\
&& \quad \times
\sqrt{\frac{(2l+1)(2l_1+1)(2l_2+1)}{4\pi}}\begin{pmatrix}l&l_1&l_2\\2&0&-2\end{pmatrix}
\, . \eea This is exactly the relation that one encounters when
lensing the polarization field by foreground density fluctuations
\cite{Hu00}.

The integral related to the curl-type displacement is tedious, but
can be simplified using relations involving raising and lowering
of the spin and gradient of the spin-weighted spherical harmonic.
To calculate \bea {}_{2}I^\otimes_{lml_1m_1l_2m_2}&=&\int d\bn
({}_{\pm2}Y_{lm}^*){}_{\pm2}Y_{l_1m_1:a}Y_{l_2m_2:b}\epsilon^{ba}
\, , \eea we note
\bea
\bm_- \cdot \nabla {}_{s}Y_{lm} &=& \sqrt{\frac{(l-s)(l+s+1)}{2}} {}_{s+1}Y_{lm} \nonumber \\
\bm_+ \cdot \nabla {}_{s}Y_{lm} &=& -\sqrt{\frac{(l+s)(l-s+1)}{2}} {}_{s-1}Y_{lm} \ ,
\eea
and the relation
\bea
(\bm_+)_i(\bm_-)_j+(\bm_-)_i(\bm_+)_j=g_{ij},
\eea
to write $\nabla\ {}_sY_{lm}=(\bm_+\bm_-+\bm_-\bm_+)\cdot\nabla {}_sY_{lm}$ as
\bea
&&{}_sY_{lm}=\\
&&\sqrt{\frac{(l-s)(l+s+1)}{2}}\
{}_{s+1}Y_{lm}\mathbf{m}_+ \nonumber \\
&&\quad \quad - \sqrt{\frac{(l+s)(l-s+1)}{2}} {}_{s-1}
Y_{lm}\mathbf{m}_- . \nonumber
\eea
This leads to
\bea
&&{}_{2}I^\otimes_{lml_1m_1l_2m_2}=\int d\bn
({}_{+2}Y_{lm}^*){}_{+2}Y_{l_1m_1:a}Y_{l_2m_2:b}\epsilon^{ba}\nonumber\\
&=&i(-1)^m\sqrt{\frac{l_2(l_2+1)}{2}}\sqrt{\frac{(2l+1)(2l_1+1)(2l_2+1)}{4\pi}}
\nonumber \\
&&\begin{pmatrix}
l&l_1&l_2\\-m&m_1&m_2\end{pmatrix}\Big(\sqrt{\frac{(l_1+2)(l_1-1)}{2}}\begin{pmatrix}
l&l_1&l_2\\2&-1&-1\end{pmatrix}\nonumber \\
&&\quad -\sqrt{\frac{(l_1-2)(l_1+3)}{2}}\begin{pmatrix}l&l_1&l_2\\2&-3&1
\end{pmatrix}\Big)\, ,\eea
such that
\begin{widetext}
\bea
{}_{2}S_1^\otimes&=&\sum_{m_1m_2}|{}_{2}I^\otimes_{lml_1m_1l_2m_2}|^2=\frac{1}{2l+1}|{}_{2}
F^\otimes_{ll_1l_2}|^2 \\
{}_{2}F^\otimes_{ll_1l_2}&=&\sqrt{\frac{l_2(l_2+1)(2l+1)(2l_1+1)(2l_2+1)}{8\pi}}\left(
\sqrt{\frac{(l_1+2)(l_1-1)}{2}}\begin{pmatrix}l&l_1&l_2\\2&-1&-1\end{pmatrix}-\sqrt{\frac{(l_1-2)(l_1+3)}{2}}
\begin{pmatrix}l&l_1&l_2\\2&-3&1\end{pmatrix}\right)\, . \nonumber
\eea
\end{widetext}
Furthermore, with $s=\pm2$, equation~(\ref{eqns}) can be
generalized to
\bea
{}_{2}S_2^\oplus=-\frac{1}{2}[l(l+1)-4]l_1(l_1+1)\frac{2l_1+1}{4\pi}
\,  , \eea and as in the case of lensed temperature  anisotropies,
\bea {}_{+2}S_2^\otimes={}_{+2}S_2^\oplus. \eea For the
cross-correlation between E-modes and the temperature, we find
\bea
{}_{02}S_1^\oplus&=&\frac{1}{2l+1}(F_{ll_1l_2}^\oplus)({}_{+2}F_{ll_1l_2}^\oplus) \nonumber \\
{}_{02}S_1^\otimes&=&\frac{1}{2l+1}(F_{ll_1l_2}^\otimes)({}_{+2}F_{ll_1l_2}^\otimes) \, .
\eea

Putting the terms together, we can write the lensed power spectra in polarization as
\begin{widetext}
\bea C_{l}^{\tilde{E}}&=&C_{l}^E - (l^2+l-4)RC_{l}^E +
\frac{1}{2(2l+1)}\sum_{l_1l_2}\left[C_{l_1}^
{h^\oplus}({}_{2}F^\oplus_{ll_1l_2})^2+C_{l_1}^{h^\otimes}({}_{2}F^\otimes_{ll_1l_2})^2\right]\left[(C_{l_2}^E+C_{l_2}^B)+(-1)^L
(C_{l_2}^E-C_{l_2}^B)\right]\nonumber \\
C_{l}^{\tilde{B}}&=&C_{l}^B - (l^2+l-4)RC_{l}^B +
\frac{1}{2(2l+1)}\sum_{l_1l_2}\left[C_{l_1}^
{h^\oplus}({}_{2}F^\oplus_{ll_1l_2})^2+C_{l_1}^{h^\otimes}({}_{2}F^\otimes_{ll_1l_2})^2\right]\left[(C_{l_2}^E+C_{l_2}^B)-(-1)^L
(C_{l_2}^E-C_{l_2}^B)\right] \nonumber \\
C_l^{\tilde{\theta}\tilde{E}}&=&C_l^{\theta E}-(l^2+l-2)RC_l^{\theta E}+\frac{1}{2l+1}\sum_{l_1l_2}\left[C_{l_1}^{h^\oplus}(F_{ll_1l_2}^\oplus)({}_{+2}F_{ll_1l_2}^\oplus) +C_{l_1}^{h^\otimes}(F_{ll_1l_2}^\otimes)({}_{+2}F_{ll_1l_2}^\otimes) \right]C_{l_2}^{\theta E} \, ,
\eea
\end{widetext}
with $R$ from equation~(\ref{eqnR}).

The case of CMB lensing by foreground density fluctuations is
simply the replacement of $C_{l_1}^{h^\oplus}$ with the power
spectrum of projected potentials and with $C_{l_1}^{h^\otimes}=0$.
The lensing contribution to the B-mode from the foreground density
field \cite{Zaldarriaga:1998ar} act as the main contaminant in
detecting the primary gravitational wave signal in B-modes of
polarization. To see the extent to which lensing by gravitational
waves themselves may become important in B-mode polarization
studies, we will assume $C_l^B$ to be zero and present a
comparison between $C_l^{\tilde B}$ from density fluctuations and
$C_l^{\tilde B}$ from gravitational waves. As we find, the
secondary lensing from gravitational waves is smaller than the
cosmic variance level of lensing B-modes from density
perturbations and will remain undetectable in anisotropy maps.
While below the cosmic variance limit, $C_l/\sqrt{l}$, for
anisotropy measurements, systematics must generally be controlled to a level
far below this; if systematics apply to a wide range of multipoles,
then one must control their effects to $C_l/l$. For lensing by foreground
gravitational waves at the maximum amplitude, we find that the corrections
are above this level when $l > 10^3$ suggesting that one only needs to be concerned of these signals
in all sky maps with no instrumental noise and other secondary signals.

\section{Results and Discussion}
\label{sec4}

In Figure~1, we show a comparison of $C_l^{h^\otimes}$,
$C_l^{h^\oplus}$, and the angular power spectrum of deflection
angle from projected density perturbations along the line of sight
to $\eta_s$ at a redshift of $1100$ corresponding to the CMB last
scattering surface. In calculating the power spectra of lensing
from foreground gravitational waves, we have assumed an amplitude
for the tensor modes with a value for $H_I$ in
equation~(\ref{eqnpk}) of $2 \times 10^{14}$ GeV. This corresponds
to a tensor-to-scalar ratio of 0.3, which is roughly the upper
limit allowed by current CMB and large-scale structure
observations \cite{WMAP3}.

The curl spectrum of deflections from gravitational waves has been
previously discussed in the literature in the context of weak
lensing surveys with galaxy shapes \cite{DRS03}. The gradient-type
displacement spectrum from gravitational waves discussed here is
also important and cannot be ignored when calculating
modifications to CMB temperature and polarization anisotropies.
Note that $C_l^{h^\otimes}$ and $C_l^{h^\oplus}$ peak at large
angular scales corresponding to $\ell=2$ to $\ell=10$. To compare
lensing by gravitational waves and lensing by mass, we calculate
the rms deflection angle through $\theta^2_{\rm rms} = [\sum
l(l+1)(2l+1)/4\pi (C_l^{h^\otimes}+C_l^{h^\oplus})]$. For
gravitational waves related spectra shown in Figure~1,
$\theta_{\rm rms}$ is $3.62 \times 10^{-5}$ radians or roughly 7
arcsecs, which is a factor of 20 smaller than the rms deflection
angle for CMB photons under density fluctuations, where
$\theta_{\rm rms} \sim 7 \times 10^{-4}$. The coherence scale,
where the rms drops to half of its peak value, is about $\sim$ 60
degrees for lensing by gravitational waves, while for density
perturbations the coherence scale is about a degree. With such a
large coherence scale and a small rms deflection angle, foreground
gravitational waves deflect large patches of the CMB sky by the
same angle of about 6 arcsecs, resulting in an overall small
modification to the CMB anisotropy and polarization spectra, when
compared to the case with density fluctuations alone.

The differences between lensing by density perturbations and
lensing by gravitational waves is clear in Figure~2, where we show
modifications to the temperature power spectrum and the cross
power spectrum between
 temperature and E-modes of polarization. The secondary
 lensing correction from the foreground gravitational waves is
smaller than the cosmic variance level of intrinsic CMB
anisotropies. Even with perfect CMB observations devoid of
instrumental noise, it is unlikely that the lensing modification
by primordial gravitational waves in the foreground of CMB will be
detectable. As shown in Figure~2, however, the corrections are above the
systematic level of $C_l/l$ for primordial anisotropy measurements when $l > 10^3$.
Thus, in the extreme case where one is dealing with perfect all-sky maps
cleaned of foregrounds and other secondary signals, one could be concerned that these
effects act as a source of systematic error for primordial anisotropy measurements.
As is clear from Figure~2, while lensing by gravitational waves is
not significant, lensing of the CMB by foreground density
perturbations will be detectable as the modifications are well
above the cosmic variance limit.

In Figure~3, we summarize our results related to lensing of
polarization anisotropies in terms of the power spectra or E- and
B-modes. Again, the lensing effect by foreground gravitational
waves is below the cosmic variance level of the dominant signal in
the E- and B-mode maps, but above the systematic level. In the case of B-modes, lensing by density
perturbations will remain the main contaminant in searching for
gravitational wave signatures from the primary B-mode power
spectrum. Lensing signals by foreground gravitational waves are
unlikely to affect reconstruction techniques that attempt to
remove the lensed B-modes when searching for a low amplitude
gravitational wave background \cite{HuOka02}.

 While we have only considered angular displacements on the sky,
 gravitational waves also contribute to a
variation along the line of sight that can be described as a
time-delay effect. Just as angular displacement couples to the
angular gradient of the CMB, the radial displacement couples to
the radial gradient of the CMB. These effects, however, are
smaller due to lack of radial structure  in the perturbations that form the
primordial anisotropy spectrum in the CMB, such as the acoustic peaks.
There are also geometric cancellations associated with the projection of line-of-sight
time-delay modulations to an anisotropy pattern on the CMB sky \cite{HuCoo01}.
Thus, it is unlikely that our conclusions related to lensing by foreground gravitational waves
are affected by including the time-delay effect.

In general, our results are consistent with those of
Ref.~\cite{DRS03} who studied the possibility of measuring the
gravitational wave background amplitude using weak lensing surveys
of galaxy shapes and using the curl-mode of the shear. Even with
an optimistic survey with a large surface density of galaxies to
measure shapes, the gravitational wave signal in the shear remains
undetectable below the noise.

To summarize our calculation, while weak lensing distortion of the
 cosmic microwave background (CMB) temperature and polarization
patterns by foreground density fluctuations is well studied in the
literature, we noted the lack of a detailed description related to
lensing modifications by foreground gravitational waves or tensor
perturbations. Here, we have presented an analytical formulation
on how CMB anisotropies and polarization patterns are distorted by
a stochastic background of primordial gravitational waves between
us and the last scattering surface. Our analytical formulation is
useful when studying general lensing of any background source by
foreground gravitational waves.

 While density fluctuations perturb CMB photons via gradient-type
 displacements only, gravitational waves distort CMB anisotropies via both the gradient- and the
curl-type displacements. The latter can be described as a rotation
of background images in the presence of foreground gravitational
waves while the former is related to the lensing convergence. For
a primordial background of gravitational waves from inflation with
an amplitude corresponding to a tensor-to-scalar ratio below the
current upper limit of $\sim$ 0.3, the resulting modifications to
the angular power spectra of CMB temperature anisotropy and
polarization are below the cosmic variance limit, but above the systematic level.
Thus, it is unlikely that planned high sensitivity CMB observations warrant an
accounting of the secondary contributions discussed here as they
are not expected to affect precise parameter measurements; if
observations are all-sky measurements with no instrumental noise, then these
effects may be present in the form of systematic corrections to the primary anisotropy and
polarization measurements.

\begin{acknowledgments}
We thank A. Stebbins and J. Pritchard for useful discussions and an anonymous
referee for useful comments. This work was supported in part by DoE at UC Irvine (AC) and by the
Moore Foundation at Caltech (CL).
\end{acknowledgments}



\begin{thebibliography}{99}

\bibitem{lensing}  See, e.g., U.~Seljak and M.~Zaldarriaga, \PRL\ {\bf
  82}, 2636 (1999) [arXiv:astro-ph/9810092]; \PRD\ {\bf 60},
  043504 (1999) [arXiv:astro-ph/9811123]; M. Zaldarriaga and
  U. Seljak, \PRD\ {\bf 59}, 123507 (1999)
  [arXiv:astro-ph/9810257]; W.~Hu, \PRD\ {\bf 64}, 083005 (2001)
  [arXiv:astro-ph/0105117].

\bibitem{Hu00}
        W.~Hu, \PRD\ {\bf 62}, 043007  (2000) [arXiv:astro-ph/0001303].


\bibitem{Cha}
  A.~Lewis and A.~Challinor,
  arXiv:astro-ph/0601594.


\bibitem{HuOka02}
     W.~Hu and T.~Okamoto, \ApJ\ {\bf 574}, 566 (2002)
     [arXiv:astro-ph/0111606];
M.~Kesden, A.~Cooray, and M.~Kamionkowski,
        \PRD\ {\bf 67}, 123507 (2003) [arXiv:astro-ph/0302536];
C.~M.~Hirata and U.~Seljak,
Phys.\ Rev.\ D {\bf 68}, 083002 (2003)
[arXiv:astro-ph/0306354].

\bibitem{Zaldarriaga:1998ar}
  M.~Zaldarriaga and U.~Seljak,
  Phys.\ Rev.\ D {\bf 58}, 023003 (1998)
  [arXiv:astro-ph/9803150].


\bibitem{KamKosSte97} M.~Kamionkowski, A.~Kosowsky, and
     A.~Stebbins, \PRL\ {\bf 78}, 2058 (1997) [arXiv:astro-ph/9609132];
        U.~Seljak and M.~Zaldarriaga, \PRL\ {\bf 78}, 2054
     (1997) [arXiv:astro-ph/9609169].


\bibitem{KesCooKam02} M.~Kesden, A.~Cooray, and M.~Kamionkowski,
     \PRL\ {\bf 89}, 011304 (2002) [arXiv:astro-ph/0202434];
     L.~Knox and Y.-S.~Song, \PRL\ {\bf 89}, 011303
     [arXiv:astro-ph/0202286]; U. Seljak and C. Hirata,
     Phys. Rev. D {\bf 69}, 043005 (2004) [arXiv:astro-ph/0310163].

\bibitem{Jaffe}
  N.~Kaiser and A.~H.~Jaffe,
  Astrophys.\ J.\  {\bf 484}, 545 (1997)
  [arXiv:astro-ph/9609043].

\bibitem{stebbins} A. Stebbins, preprint, arXiv:astro-ph/9609149.

\bibitem {DRS03}
 S.~Dodelson, E.~Rozo and A.~Stebbins,
  Phys.\ Rev.\ Lett.\  {\bf 91}, 021301 (2003)
  [arXiv:astro-ph/0301177].

\bibitem{CooKamCal05}
  A.~Cooray, M.~Kamionkowski and R.~R.~Caldwell,
  Phys.\ Rev.\ D {\bf 71}, 123527 (2005)
  [arXiv:astro-ph/0503002].


\bibitem{CooHu02}
A.~Cooray and W.~Hu,
Astrophys.\ J.\  {\bf 574}, 19 (2002)
[arXiv:astro-ph/0202411];
C.~Shapiro and A.~Cooray,
  arXiv:astro-ph/0601226.
  C.~M.~Hirata and U.~Seljak,
  Phys.\ Rev.\ D {\bf 68}, 083002 (2003)
  [arXiv:astro-ph/0306354].
\bibitem{Gallens}
J.~Miralda-Escude, Astrophys.\ J.\  {\bf 380}, 1 (1991); R.
~D.~Blandford, A.~B.~Saust, T. G.~Brainerd and J.~Villumsen, Mon.\
Not.\ Roy.\ Astron.\ Soc.\ {\bf 251}, 600 (1991); N.~Kaiser,
Astrophys.\ J.\  {\bf 388}, 272 (1992); For recent reviews, see,
M.~Bartelmann and P.~Schneider, Phys. Rept. {\bf 340}, 291 (2001);
P.~Schneider  Gravitational Lensing: Strong, Weak \& Micro,
Lecture Notes of the 33rd Saas-Fee Advanced Course, (Berlin:
Springer-Verlag)

\bibitem{timedelay}
I.~I.~Shapiro, Phys. Rev. Lett, {\bf 13} 789 (1964)


\bibitem{HuCoo01}
  W.~Hu and A.~Cooray,
  Phys.\ Rev.\ D {\bf 63}, 023504 (2001)
  [arXiv:astro-ph/0008001].

\bibitem{Prit}
  J.~R.~Pritchard and M.~Kamionkowski,
  Annals Phys.\  {\bf 318}, 2 (2005)
  [arXiv:astro-ph/0412581].

\bibitem{Gol67}
J.~N.~Goldberg, et al. J. Math. Phys., 7, {\bf 863} (1967).

\bibitem{spin1}
A.~Challinor and G.~Chon,
  Phys.\ Rev.\ D {\bf 66}, 127301 (2002)
  [arXiv:astro-ph/0301064].


\bibitem{spin2}
T.~Okamoto and W.~Hu,
  Phys.\ Rev.\ D {\bf 67}, 083002 (2003)
  [arXiv:astro-ph/0301031];


\bibitem{WMAP3} D.~N.~Spergel,et.al, preprint, arXiv:astro-ph/0603449.

\end{thebibliography}
\end{document}